# Formation of plano-convex micro-lens array in fused silica glass using $CO_2$ laser assisted reshaping technique

Ik-Bu Sohn,[1,*] Hun-Kook Choi,[1,2] Dongyoon Yoo,[1] Young-Chul Noh,[1] Md. Shamim Ahsan,[3,**] Jae-Hee Sung,[1] and Seong-Ku Lee[1]

[1]*Advanced Photonics Research Institute (APRI), Gwangju Institute of Science and Technology (GIST), 1 Oryong-dong, Buk-gu, Gwangju 61005, Republic of Korea*

[2]*Department of Photonic Engineering, Chosun University, 309 Pilmun-daero, Dong-gu, Gwangju 61452, Republic of Korea*

[3]*Electronics and Communication Engineering Discipline, School of Science, Engineering, and Technology, Khulna University, Khulna 9208, Bangladesh*

We report on fabricating high-fill-factor plano-convex spherical and square micro-lens arrays on fused silica glass surface using $CO_2$ laser assisted reshaping technique. Initially, periodic micro-pillars have been encoded on the glass surface by means of a femtosecond laser beam. Afterwards, the micro-pillars are polished several times by irradiating a $CO_2$ laser beam on top of the micro-pillars. Consequently, spherical micro-lens array with micro-lens size of 50 μm × 50 μm and square micro-lens array with micro-lens size of 100 μm × 100 μm are formed on fused silica glass surface. We also study the intensity distribution of light passed through the spherical micro-lens array engraved glass sample. The simulation result shows that, the focal length of the spherical micro-lens array is 35 μm. Furthermore, we investigate the optical properties of the micro-lens array engraved glass samples. The proposed $CO_2$ laser based reshaping technique is simple and fast that shows promises in fabrication arrays of smooth micro-lenses in various transparent materials.

Email: *ibson@gist.ac.kr, **shamim@kaist.ac.kr

Fax: +82-(0)62-715-3445

## I. INTRODUCTION

Direct fabrication of micro/nano-metric structures on various materials is of significant importance in manufacturing various functional devices. Currently, a large variety of patterning technologies exist in the industry for precise patterning of transparent and non-transparent materials [1-6]. Still the industry is in need of simple and flexible single-step patterning technology for micro-machining of various materials easily at lowered processing cost. Since the discovery of the first functional coherent laser in 1960, lasers have proved their importance to pattern materials in micro/nano scale. During last several decades, a good number of laser systems have been widely adopted for micro/nano-structuring of materials: nanosecond laser [7], picosecond laser [8,9], femtosecond laser [10-15], solid state laser [16], and $CO_2$ laser [17-20]. $CO_2$ lasers have been serving as one of the most useful laser sources after its invention in 1964 for micro-fabrication and melting of various substrates [17-20]. Several research groups conducted researches utilizing high power $CO_2$ laser sources. The researchers investigated the opportunity of drilling alumina using high peak power $CO_2$ laser system [17]. The enormous potential of $CO_2$ laser beam for rapid micro-structuring of polymers to develop microfluidic systems was reported by another research group [18]. Researchers also reported an advanced $CO_2$ laser processing technology with poly-dimethylsiloxane (PDMS) protection for crack/defect-less micro-machining of glass materials [19]. $CO_2$ laser was also utilized for surface polishing of glass materials in a large area, where surface roughness up to 500 nm was efficiently lowered with final rms values around 1 nm [20]. Although $CO_2$ laser is suitable for wide application areas, but fabrication of micro-lenses using $CO_2$ laser is very much challenging.

There have been reports on the formation of micro-lenses on the surface of various materials using diverse technologies [21-28]. One Research group fabricated closely-joined convex micro-lens array (MLA) on glass surface [21]. Initially, periodic micro-holes, i.e., concave MLAs were printed on glass surfaces. These samples were then experienced chemical etching to pattern circular or rectangular shape structures. Using hot embossing, this concave shape periodic pattern was transferred on polymethyl methacrylate (PMMA) surface that develop convex MLAs on the PMMA surface. Using femtosecond laser systems, the same research group proposed the formation of concave MLAs on cylindrical glass [22] and polydimethylsiloxane (PDMS) [23]. Another research group proposed the formation of concave MLA in glass using femtosecond laser direct writing followed by thermal treatment, wet etching, and additional annealing [24]. Femtosecond laser beam was irradiated inside glass using a complex patterning software to form cylindrical and bump shape structures inside glass samples. After heat treatment and chemical etching, cylindrical and spherical shape MLAs were developed on glass surface. Researchers reported the fabrication of micro-lens arrays in polycarbonate with nano-joule energy femtosecond laser by simply focusing the laser beam inside the polycarbonate [25]. Heat accumulation at the laser focal point and subsequent material expansion leads to localized swelling on the sample surface that acts as a micro-lens. Spherical and square shape MLAs have been reported by another group of researchers by means of femtosecond laser direct writing and subsequent baking & chemical etching [26]. There is evidence of forming concave MLAs on glass plate using $CO_2$ laser beam [27]. Rod and spherical shape silica micro-lenses were fabricated by melting glass cylinder of appropriate size and shape using $CO_2$ laser beam [28]. Most of these technologies are somewhat complex requiring multi-stage processing. Besides, MLA fabrication over a large sample area with consistent size and shape is quiet difficult. As a result, it is still challenging to develop plano-convex spherical and square micro-lens arrays of flexible size and shape over a large sample area. In one of our previous works, we reported the formation of cylindrical MLAs on fused silica glass by means of laser irradiations [29].

In this paper, we demonstrate a novel technique to fabricate plano-convex spherical and square micro-lens arrays on fused silica glass surface by making use of the melting capability of $CO_2$ laser sources. The key structure responsible for the formation of plano-convex MLAs, is the periodic micro-pillars. Initially, we printed micro-gratings of suitable size and period on fused silica glass surface using femtosecond laser direct writing. Using the same technique, micro-gratings of same size and period have been printed at right angles with the previously printed micro-gratings. As a consequence, periodic micro-pillars are formed in between the micro-grooves. Afterwards, $CO_2$ pulsed laser beam has been exposed in both the horizontal and vertical directions on top of the micro-pillars engraved glass surface. Plano-convex MLAs are evident after single time $CO_2$ laser beam irradiation. Subsequent $CO_2$ laser beam exposures cause the smoothening of the plano-convex MLAs. Furthermore, we simulate the intensity distribution of light at different positions after the spherical MLA engraved glass sample, which in turn gives us the focal length of the spherical micro-lenses. We also investigate the optical properties of the MLA encoded glass samples. The proposed technique is suitable for large area fabrication of MLAs and diffractive optical elements (DOEs).

## II. EXPERIMENTAL DETAILS

### 2.1 Laser Systems

We fabricated periodic micro-pillars on fused silica glass surface using a Ti:sapphire femtosecond laser system (IFRIT, Cyber laser) operating at the central wavelength of 786 nm that emits ultrashort laser pulses of 250 fs pulse width at a pulse repetition rate ($R_p$) of 1kHz. Highly transparent fused silica glass samples (transparency: greater than 90% in the visible spectrum; thickness: 1 mm) with refractive index of 1.458 at 588 nm were used during our experiments. S-polarized Gaussian beam, obtained after the polarizing beam splitter, was delivered through a beam expander (LiNOS; Magnification: ~ 2×) to expand the laser beam. The laser samples were placed on a motorized 3-axis linear translation stage, the resolution of which is 100 nm in all directions. The femtosecond laser beam was focused on the fused silica glass surface using an achromatic objective lens having magnification of 10× (Mitutoyo; M Plan Apo NIR; numerical aperture: 0.26). Initially, periodic micro-gratings, i.e. micro-grooves of desired size and period were engraved on the glass surface. Afterwards, micro-grooves of similar size and period were fabricated at right angles to the previously encoded micro-grooves. As a consequence, micro-pillars array has been developed throughout the sample glass. The schematic diagram of the femtosecond laser system is illustrated in Fig. 1(a).

A $CO_2$ pulsed laser system (Coherent, C-55L) operating at the wavelength of 10.6 μm with Rp of 5 kHz was used to reshape the micro-pillars into micro-lenses. The pulse width of the laser beam was 120 ± 40 μs. The micro-pillars array engraved glass samples, patterned by femtosecond laser writing, were placed on a 3-axis translation stage having resolution of 100 nm in all directions. We focused the attenuated $CO_2$ laser beam of moderate laser energy on top of the micro-pillars array by means of a galvanometer scanner (LiNOS; focal length: 170 mm), which was irradiated several times in both X and Y directions of the sample. Fig. 1(b) depicts the schematic diagram of the $CO_2$ laser polishing system. The fabrication process of plano-convex MLA is illustrated in Fig. 1(c).

## 2.2 Measurements and Analysis

We analyzed the surface morphology of the plano-convex MLA on fused silica glass surface by means of a field emission scanning electron microscope (FE-SEM) (Hitachi, S-4700). Before SEM examination, the glass samples were sputter-coated by ~ 30 nm layer of gold coating to provide conductivity for SEM examination. To investigate the profile image of the micro-gratings' engraved glass samples, we utilized a confocal microscope (Olympus, OLS3100). We further examined the plano-convex MLA under an optical microscope (ZEISS, Axioskop 40). We also measured the transmittance and absorbance of light for both the spherical and square MLA engraved glass samples using Spectrometer (Jasco Corp., V-570) in the wavelength range of 200-900 nm. The intensity distribution of light at different positions after the spherical MLA engraved glass sample was simulated using LightTools simulator, from which we obtained the focal length of the spherical micro-lenses. During simulation, we considered the following simulation parameters.

- Radius of Curvature (R): 20.5 μm
- Height ($H$): 16 μm
- Micro-lens Array: 10 ×10 micro-lenses
- Material: Fused silica glass (Thickness: 1 mm; Refractive index: 1.458)
- Wavelength of Light: 632 nm

## III. RESULTS

We fabricated both spherical and square shape plano-convex MLAs on fused silica glass surface using $CO_2$ laser assisted reshaping technique. The key structure for forming plano-convex MLAs is the micro-pillars array, which has been encoded on the glass surface by means of single beam femtosecond laser writing. $CO_2$ laser assisted polishing converts the micro-pillars to plano-convex MLAs. The type of the MLAs is dependent on the area, depth, and period of the micro-pillars and associated irradiation parameters of the $CO_2$ laser beam. The details of the experimental results are discussed briefly in the following sub-sections.

### 3.1 Formation of plano-convex Spherical Micro-lens Array

The fabrication technique of plano-convex spherical MLA on glass surface involves two fundamental processing steps. The first step is to encode periodic micro-pillars on the glass surface by irradiating a femtosecond laser beam of 24 J/cm$^2$ laser fluence five times (in both the x and y directions) at a scanning speed of 100 mm/s and a scanning step of 60 μm. The width and period of the linear gratings, engraved at right angles, were 10 μm and 60 μm. As a consequence, periodic micro-pillars of 50 μm × 50 μm area were formed throughout the sample area. Fig. 2(a) illustrates the optical microscope (OM) image of the periodic micro-pillars' array developed on the sample surface. The 3D confocal microscope (CM) image of the periodic micro-pillars' array is depicted in Fig. 2(b). We have also investigated the profile of the micro-gratings, the alpha step profile image of which is represented in Fig. 2(c). The second step of forming plano-convex spherical MLA is to reshape the femtosecond laser engraved micro-pillars array by means of a $CO_2$ laser beam. A $CO_2$ laser beam of 8.4 J/cm$^2$ laser fluence was irradiated in both X and Y directions just on top of the micro-pillars at a scanning speed of 20 mm/s and a scanning step of 20 μm. Single time $CO_2$ laser irradiation accomplished plano-convex spherical micro-lens of 50 μm × 50 μm area throughout the glass surface. Fig. 2(d) shows the optical microscope image of the plano-convex spherical MLA where the period remains 60 μm. The SEM image (top view) of the 60 μm period spherical MLA is shown in Fig. 2(e). The magnified SEM image of Fig. 2(e) is illustrated in Fig. 2(f); the side view of the spherical MLA is shown in Fig. 2(g). The radius of curvature (R) and the height (H) of the spherical micro-lenses formed after single time $CO_2$ laser irradiation was 20.5 μm and 16 μm. Further increase in the number of $CO_2$ laser irradiation caused the smoothening of the spherical micro-lenses. The top view of the spherical MLA after 5-times $CO_2$ laser irradiation and the side view of the spherical MLA after 7-times $CO_2$ laser irradiation are depicted in Figs. 2(h) and 2(i). The radius of curvature increased with the number of $CO_2$ laser irradiation although the height of the micro-lenses from the glass surface decreased. The radius of curvature and height of the micro-lenses of Figs. 2(h) and 2(i) are 27 μm & 14.5 μm and 28 μm & 14 μm. The spherical MLA was fabricated over an area of 1 cm × 1cm. it took ~ 6 minutes for patterning linear gratings on the glass surface; whereas single time $CO_2$ laser polishing required ~ 10 minutes. Thus, the fabrication time of spherical MLA over 1 cm × 1 cm area using 1-time $CO_2$ laser polishing was ~ 16 minutes.

We also simulated the intensity distribution of light at different positions (35 μm, 44.85 μm, 55.1 μm, 80 μm, 100 μm, and 150 μm) after the plano-convex spherical MLA engraved glass sample using LightTools simulator. The highest intensity of light was observed at 35 μm from the glass sample, indicating focal point of the micro-lenses. The circular spots of Fig. 3(a) indicate obvious focusing of the spherical micro-lenses at 35 μm. The defocusing of light started after the focal point of the incident light beam, which increases with the distance from the glass sample (Figs. 3(b-f)). Consequently, the intensity of light decreased with the distance from the glass sample.

### 3.2 Formation of plano-convex Square Micro-lens Array

To accomplish plano-convex square MLA with size of 100 μm × 100 μm, we imprinted periodic micro-gratings' array in both the X and Y directions by five times repetitive application of a femtosecond laser beam of 24 J/cm$^2$ laser fluence at a scanning speed of 100 mm/s and a scanning step of 110 μm on the glass surface. Due to femtosecond laser irradiation, periodic micro-gratings with gratings' period of 110 μm and gratings' width of ~ 10 μm were encoded on the sample glass. Consequently, periodic micro-pillars of 100 μm × 100 μm area were developed on the glass sample. The optical microscope image of the micro-pillars' array is illustrated in Fig. 4(a), whereas the confocal microscope image is depicted in Fig. 4(b). The alpha step profile image of the micro-gratings is shown in Fig. 4(c).

Afterwards, the micro-pillars' engraved glass sample was polished by a $CO_2$ laser beam of 8.4 J/cm$^2$ laser fluence at a scanning speed of 20 mm/s and a scanning step of 20 μm. The $CO_2$ laser beam was focused on top of the micro-pillars and irradiated in both the X and Y directions, which initiated melting on the surface and side walls of the micro-pillars. Single time $CO_2$ laser irradiation was enough to form 110 μm period plano-convex square MLA of 100 μm × 100 μm area throughout the sample area. The optical microscope image and SEM image of the square MLA, formed after 1-time $CO_2$ laser polishing, are illustrated in Fig. 4(d) and 4(e). Further increase in the number of $CO_2$ laser irradiation caused the smoothening of the square MLA although the shape remained unchanged. The side view of the plano-convex square MLAs after several times $CO_2$ laser irradiation is illustrated in Figs. 4(f-i). The fabrication time required for forming square MLA using 1-time $CO_2$ laser polishing over an area of 1 cm × 1cm was ~ 13 minutes (~ 3 minutes for patterning linear gratings and ~ 10 minutes for 1-time $CO_2$ laser polishing).

**3.3 Optical Properties of the Spherical and Square Shape MLA Engraved Glass Samples**

We also investigated the optical properties of the MLA engraved glass samples in the visible spectrum. Fig. 5 represents the transmittance and absorbance of light of the MLA encoded (both spherical and square MLAs) glass samples, formed after several times $CO_2$ laser polishing. Experimental results confirm significant reduction of transparency of the MLA engraved glass samples compared to bare fused silica glass sample. On the contrary, absorbance of light has increased due to the formation of MLAs on the glass surface. The unpolished glass samples containing periodic gratings (Figs. 2(a) and 4(a)) show lowest transmittance (highest absorbance). Increase in the number of $CO_2$ laser polishing caused the increase in the transmittance (decrease in the absorbance) of light. The highest transmittance (lowest absorbance) was observed after 3-times $CO_2$ laser polishing; beyond which the transmittance of light started to decrease (absorbance started to increase) again.

**IV. FORMATION MECHANISM OF MICRO-LENS ARRAY ON GLASS SURFACE**

One of the most significant properties of $CO_2$ laser sources is their melting capability of transparent materials, especially, glasses. The melting of glass materials is a complex combination of photo-chemical and photo-thermal interaction. Some chemical bonds of macro-molecules are broken because of chemical reaction caused by photon absorption, while other chemical bonds are broken thermally through the released heat of the excited molecules due to photon absorption. At high laser fluence, photo-thermal interaction dominates over photo-chemical interaction [30]. Since the laser fluence of the $CO_2$ laser beam during our experiments was 8.4 J/cm$^2$, we can expect the domination of photo-thermal interaction of the $CO_2$ laser beam with the glass molecules. When a focused $CO_2$ laser beam hits the glass surface, the temperature of the irradiated spot rises rapidly that causes the melting of the glass material. Due to the exposure of the $CO_2$ laser beam over the gratings' engraved glass surface, a pool of molten glass develops around the focal point of the $CO_2$ laser beam. Afterwards, the melted material is re-solidified in the wake of the $CO_2$ laser beam [19]. The spot size of the $CO_2$ laser beam is ~ 100 μm at the focal point, which is higher than the period of the micro-lenses in both cases. Consequently, the micro-pillars formed in between the periodic grooves starts to melt. The surface tension of the molten glass, formed on top of the micro-pillars and the side walls due to $CO_2$ laser irradiation, is responsible for the curvature of the micro-pillars before re-solidification of the molten glass [28,31], which in turn causes the formation of MLAs. Although single time $CO_2$ laser irradiation is enough to form spherical and square MLA on the glass surface, repetitive application of $CO_2$ laser beam smoothens the MLAs.

## V. CONCLUSION

In summary, we fabricated plano-convex spherical and square MLAs on fused silica glass surface by polishing the femtosecond laser engraved periodic micro-pillars array using $CO_2$ laser beam. Spherical MLA with micro-lens size of 50 μm × 50 μm and square MLA with micro-lens size of 100 μm × 100 μm were developed on fused silica glass surface over 1 cm × 1 cm area. From simulation result we can infer that, the focal length of the spherical MLA is about 35 μm. The $CO_2$ laser beam causes the melting of the glass surface. The surface tension of the molten glass and their re-solidification is the primary factor responsible for the curvature of the MLAs. The experimental results confirm significant reduction in the transmittance (increase in the absorbance) of the MLA engraved glass samples. The proposed $CO_2$ laser assisted reshaping technique is fast, reliable, repeatable, and flexible showing promises for commercial production of large variety of micro-lenses in transparent materials.

ACKNOWLEDGEMENTS

This work was supported by the "Asian Laser Center Program" through a grant proviede by the Gwanju Institute of Science and Technology in 2015. This work was supported by the ICT R&D program of MSIP/IITP. [R0146-15-1003, Infrastructure for femto-technology]

[28] S. Calixto, M. R. Aguilar, F. J. S. Martin, L. C. Escobar, "Rod and spherical silica microlenses fabricated by $CO_2$ laser melting," Appl. Opt. **44**, 4547-4556 (2005).

[29] H. -K. Choi, M. S. Ahsan, D. Yoo, I. -B. Sohn, Y. -C. Noh, J. T. Kim, D. Jung, J. H. Kim, "Formation of cylindrical micro-lens array in fused silica glass using laser irradiations," Proc. SPIE **8923**, 89234T (2013).

[30] D. Bäuerele, *Laser Processing and Chemistry* (Springer, 2011), Chap. 10.

[31] U. C. Paek, A. L. Weaver, "Formation of s spherical lens at optical fiber ends with a $CO_2$ laser," Appl. Opt. **14**, 294-298 (1975).

Figures.

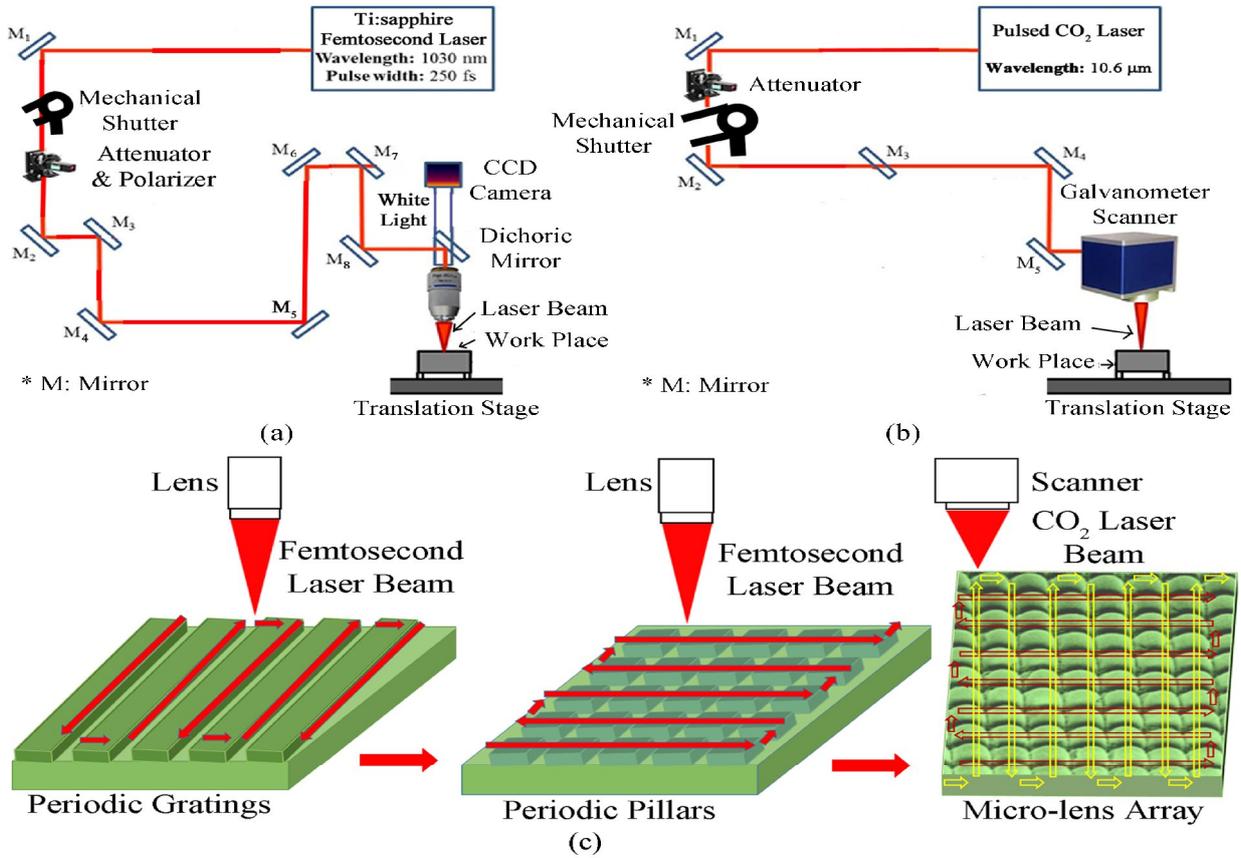

Fig. 1. Experimental setup. (a) Femtosecond laser system for patterning micro-pillars array; (b) $CO_2$ laser system for fabricating plano-convex micro-lens array; (c) fabrication process of plano-convex micro-lens array.

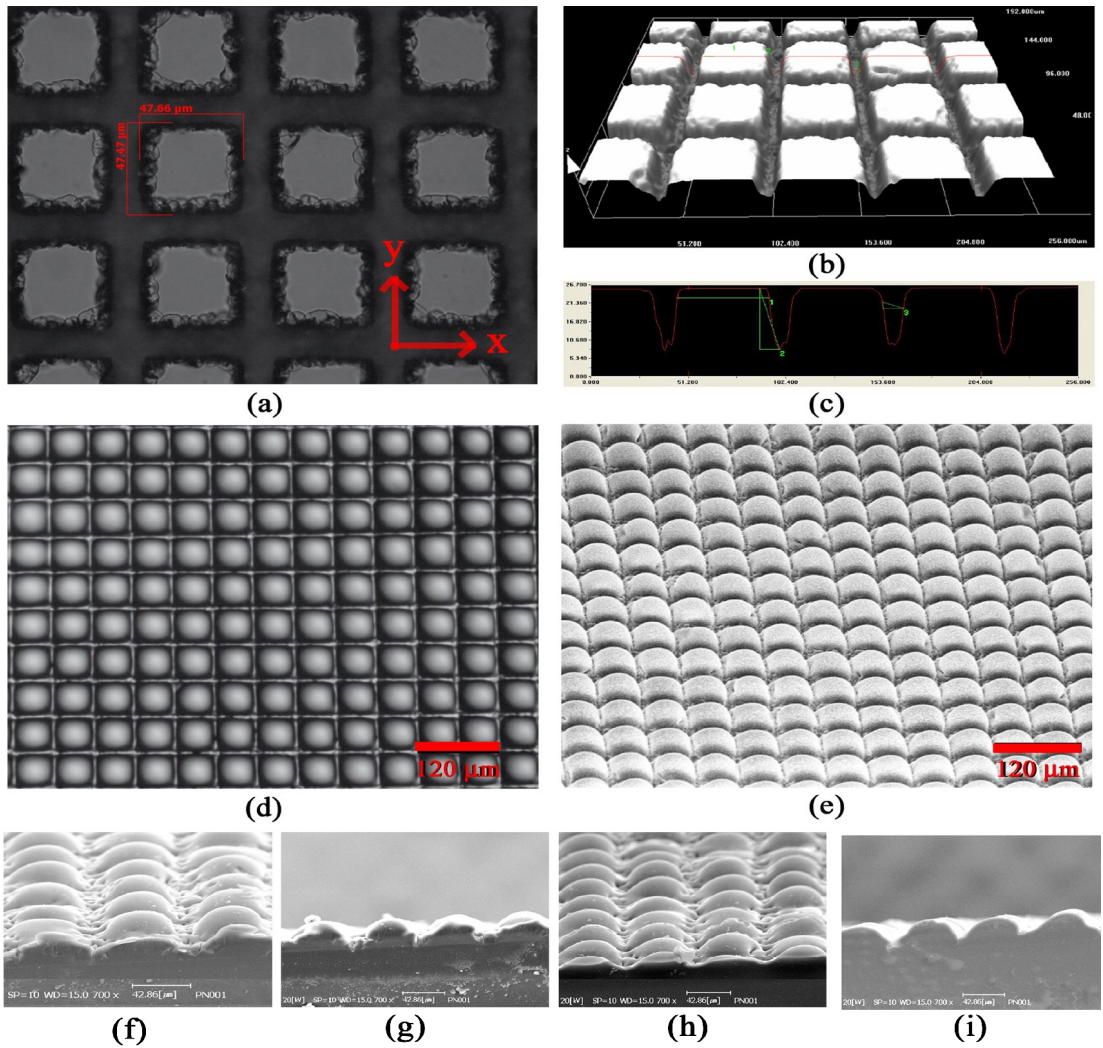

Fig. 2. Formation of 60 μm period plano-convex spherical MLA with micro-lens size of 50 μm × 50 μm on fused silica glass surface using $CO_2$ laser assisted reshaping technique. (a) OM image of the femtosecond laser engraved micro-pillars' array before laser polishing; (b) 3D CM image of the micro-pillars' array of Fig. 2(a); (c) alpha step profile image of the micro-gratings of Figs. 2(a) and 2(b); (d) OM image of the spherical MLA after 1-time $CO_2$ laser polishing; (e) SEM image (top view) of the spherical MLA of Fig. 2(d); (f,g) magnified SEM image (top view and side view) of the spherical MLA after 1-time $CO_2$ laser polishing; (h) SEM image (top view) of the MLA after 5-times $CO_2$ laser polishing; (i) side view of the MLA after 7-times $CO_2$ laser polishing.

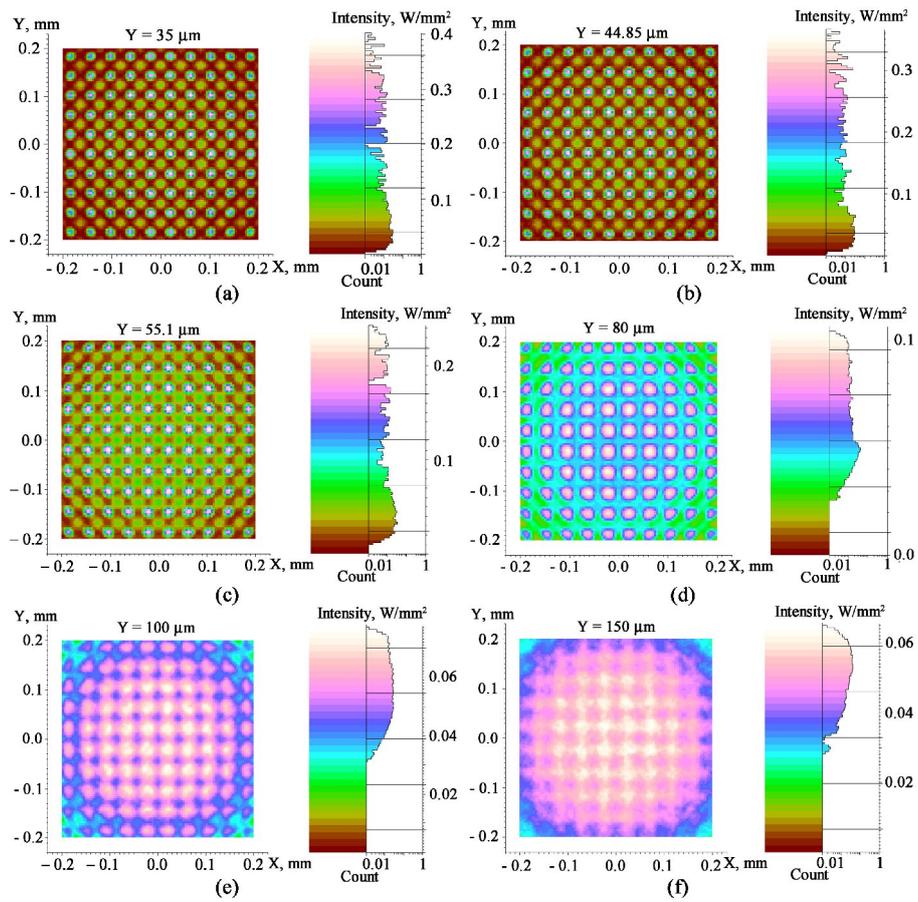

Fig. 3. Intensity distribution of light at different positions after the plano-convex spherical MLA engraved glass sample. After (a) 35 μm; (b) 44.85 μm; (c) 55.1 μm; (d) 80 μm; (e) 100 μm; (f) 150 μm.

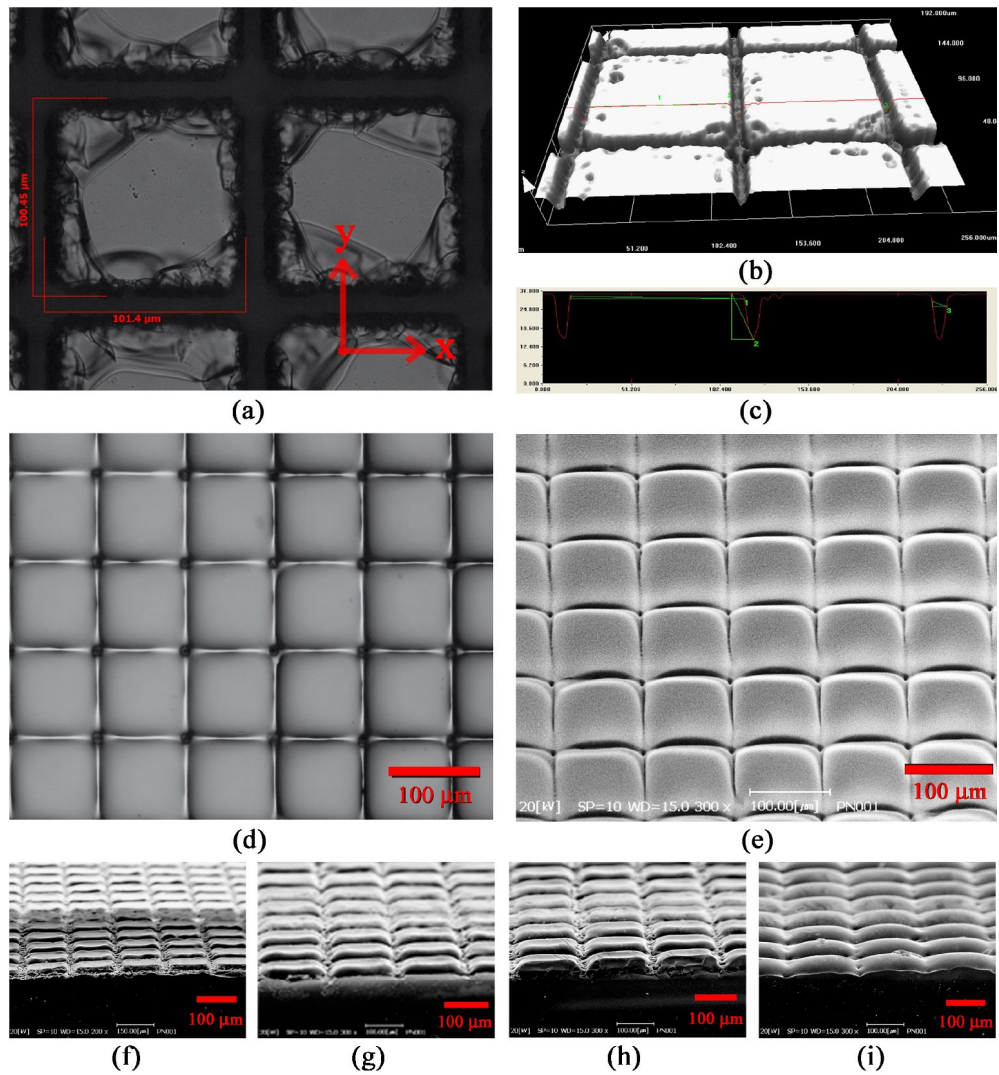

Fig. 4. Formation of plano-convex square MLA of 100 μm×100 μm area on fused silica glass surface using CO$_2$ laser assisted reshaping technique. (a) OM image of the femtosecond laser engraved micro-pillars' array before laser polishing; (b) 3D CM image of the micro-pillars' array of Fig. 4(a); (c) alpha step profile image of Figs. 4(a) and (b); (d) OM image of the square MLA after 1-time CO$_2$ laser polishing; (e) SEM image of the square MLA of Fig. 4(d); (f-i) side view of the square MLA: (f) 1-time polishing; (g) 3-times polishing; (h) 5-times polishing; (i) 7-times polishing.

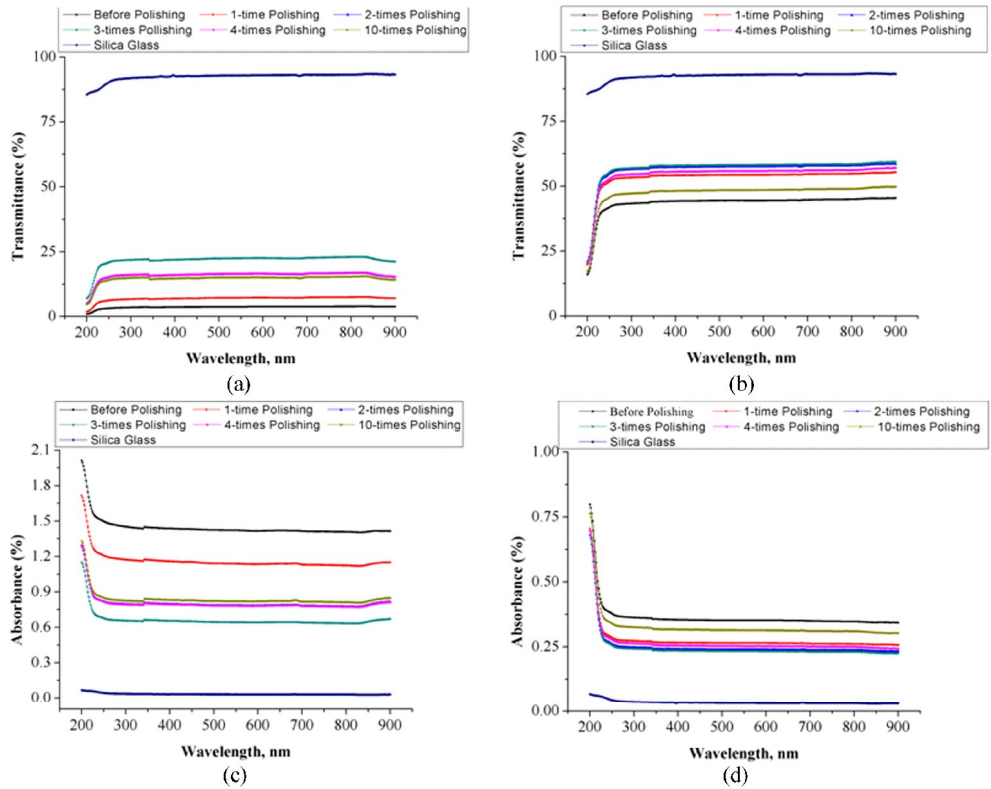

Fig. 5. Optical properties of the spherical and square shape MLA engraved glass samples. (a,c) Transmittance and absorbance of light passed through the spherical MLA engraved glass samples of Fig. 2; (b,d) Transmittance and absorbance of light passed through the square MLA engraved glass samples of Fig. 4.